\documentstyle[pra,aps]{revtex}

\begin{document}
\bibliographystyle{prsty}
\draft

\title{\bf Complete solution of the Schr\"odinger equation for the time-dependent linear potential}
\author{Mang Feng
\thanks{Electronic address: feng@mpipks-dresden.mpg.de}} 
\address{$^{1}$Max Planck Institute for the Physics of Complex Systems,\\
N$\ddot{o}$thnitzer Street 38, D-01187 Dresden, Germany\\
$^{2}$Laboratory of Magnetic Resonance and Atomic and Molecular Physics,\\ 
Wuhan Institute of Physics and Mathematics, Academia Sinica,\\
Wuhan, 430071, People's Republic of China}
\date{\today}
\maketitle

\begin{abstract}

The complete solutions of the Schr\"odinger equation for a particle with time-dependent mass moving 
in a time-dependent linear potential are presented. 
One solution is based on the wave function of the plane wave, and the other is with the form
of the Airy function. A comparison is made between the present solution and former ones to show the 
completeness of the present solution.
\end{abstract}
\vskip 0.1cm

The analytical solution of the Schr\"odinger equation with explicitly time-dependent potential has drawn much attention over
past decades. Besides the intrinsic mathematical interest, this problem connects with various applications to many
physical problems, for example, the degenerate parametric amplifier ${[1]}$ and
the quantum motion of trapped ions in the Paul 
trap ${[2]}$. To make clear the dynamical properties of the system with explicitly time-dependent potential, 
the numerical simulation can be generally applied. However, some information about the system, such as the Berry 
phase ${[3]}$ and the squeezing 
property ${[4]}$ will be probably neglected unless we can obtain the completely analytical solution of the system.
 
Not all systems with explicitly time-dependent potentials can be solved analytically. During the past several
years, some efforts have been invested in finding the solution of the time-dependent harmonic oscillator (TDHO) Hamiltonian. The most 
famous work in this respect is the invariant approach proposed by Lewis and Riesenfeld ${[5]}$. In terms of this idea 
and other elaborate methods, 
the TDHO Hamiltonian has been investigated from different angles and for different physical problems ${[6]}$. As far as we know, the 
general TDHO Hamiltonian with the potential of $g_{2}(t)x^{2}+g_{1}(t)x+g_{0}(t)$, where $g_{i}(t)$ (i=0,1,2) are arbitrary
time-dependent variables, has been solved, and the exact but very complicated form of the 
corresponding wave function  has been presented ${[7]}$. Recently, a more general TDHO problem was studied, in which the exact form 
of the propagator could be found ${[8]}$. Moreover, the investigation in this respect has been extended to the TDHO Hamiltonian with 
additional potentials ${[9,10]}$, where the result in Ref.[9] can be used to approximately describe the dynamics of two trapped 
cold ions in the Paul trap ${[11]}$.

Besides the TDHO problem, the linear potential model has also been frequently employed in some other studies ${[12,13]}$. 
Recently, this model was investigated quantum mechanically ${[14]}$, in which an analytical wave function solution for such a system 
was presented by means 
of the invariant method. Although the author of [14] claimed that his result is the first presentation in this respect, 
such a problem has actually been studied before ${[15]}$, in which the solution with 
the form of the Airy function was presented. It was shown ${[15,16]}$ that the solution with the Airy function for describing the
behavior of the free particle corresponds to a wave packet moving 
acceleratively with no change of form. However, the acceleration of the Airy packet is not the behavior of any individual particle, but
the caustic of the family of particle orbits. So there is no contradiction with Ehrenfest's theorem that no wave packet can accelerate 
in free space.

The purpose of the present paper is to undertake a completely analytical solution for the problem above along the idea in [15,16] by
means of a simple algebra, named 'time-space transformation method' ${[17]}$. With the time-space transformation 
method,  in [17], we transformed the Schr\"odinger equation with TDHO into that with time independent harmonic oscillator. But here we 
will try to transform a Schr\"odinger equation with time-dependent linear potential into that of a free particle. According to [15,16], 
there are only two solutions with nonspreading properties for the quantum treatment of a free particle. One solution is based on the 
wave function of the plane wave, and the other is with the form of the Airy function. However, as far as we know, no one has reported
these two solutions simultaneously in treating the Hamiltonian with time-dependent linear potential. Therefore, 
in what follows, we will consider a more general case 
than in Ref.[14,15], i.e., a particle with time-dependent mass moving in the time-dependent linear potential. It can be found that 
the solution in Ref.[14] is merely a particular case for a 'standing' particle under the potential, in comparison 
with our result. Besides, we will present specifically the analytical solution of this problem with the form of the Airy function. 

Consider the Schr\"odinger equation for a particle with time-dependent mass
moving in a time-dependent linear potential, which can be described by the Schr\"odinger equation in the unit of $\hbar=1$,
\begin{equation}
i\frac {\partial}{\partial t}\Psi(x,t) = -\frac {1}{2M(t)} \frac {\partial^{2}}{\partial x^{2}}\Psi(x,t) +
g_{1}(t)x\Psi(x,t)
\end{equation}
where $M(t)$ and $g_{1}(t)$ are arbitrary time-dependent variables. Performing a unitary transformation 
$\Psi(x,t)=\Phi(x,t)e^{i\beta(t)x}$ with $\beta(t)$ being a time-dependent variable determined later, we have 
$$ i\frac {\partial}{\partial t}\Phi(x,t) = -\frac {1}{2M(t)} \frac {\partial^{2}}{\partial x^{2}}\Phi(x,t) -
i\frac {\beta(t)}{M(t)}\frac {\partial}{\partial x}\Phi(x,t) + $$
\begin{equation}
\frac {\beta(t)^{2}}{2M(t)}\Phi(x,t)+ x\dot{\beta}(t)\Phi(x,t) + g_{1}(t)x\Phi(x,t)
\end{equation}
where the dot on the variable denotes the derivative with respect to time. If we perform the time and space transformation
of $y=x+\nu(t)$ and $s=\int_{0}^{t} \frac {d\sigma}{M(\sigma)}$, where $\nu(t)$ will be determined later, Eq.(2) is changed to
$$\frac {i}{M(t)}\frac {\partial}{\partial s} f(y,s) +i\dot{\nu}(t) \frac {\partial}{\partial y} f(y,s) = -\frac {1}{2M(t)} 
\frac {\partial^{2}}{\partial y^{2}} f(y,s) -i\frac {\beta(t)}{M(t)}\frac {\partial}{\partial y} f(y,s) + $$
\begin{equation}
[g_{1}(t)+\dot{\beta}(t)][y-\nu(t)]f(y,s) + \frac {\beta(t)^{2}}{2M(t)} f(y,s)
\end{equation}
in which $\Phi(x,t)=f(y,s)$ is used. To delete the term of $\frac {\partial}{\partial y} f(y,s)$, we set 
$\dot{\nu}(t)=-\frac {\beta(t)}{M(t)}$. Thus
\begin{equation}
\frac {i}{M(t)}\frac {\partial}{\partial s} f(y,s) = -\frac {1}{2M(t)} \frac {\partial^{2}}{\partial y^{2}} f(y,s) 
 + [g_{1}(t)+\dot{\beta}(t)]y f(y,s) + G(t) f(y,s) 
\end{equation}
where $G(t)= \frac {\beta^{2}(t)}{2M(t)}-[g_{1}(t)+\dot{\beta}(t)]\nu(t)$. If we assume $g_{1}(t)+\dot{\beta}(t)=0$, and 
$f^{'}(y,s) = f(y,s)e^{-i\int_{0}^{t}G(t')dt'}$, we will obtain following equation for a free particle with mass equivalent to 1
\begin{equation}
i\frac {\partial}{\partial s} f^{'}(y,s) = -\frac {1}{2} \frac {\partial^{2}}{\partial y^{2}} f^{'}(y,s) 
\end{equation}
From the usual textbook of quantum mechanics, we know that the simplest form of the solution is 
$f^{'}(y,s)=\frac {1}{\sqrt{2\pi}} e^{i(Ay- A^{2}s/2)}$ with 
$A$ being an arbitrary real number if we define the particle propagating or counter-propagating along the direction of $y$.
Reversing the procedure above, we can obtain
\begin{equation}
\Psi (x,t) = \frac {1}{\sqrt{2\pi}} \exp\{iA[x+\nu(t)]\} \exp\{-i \frac {A^{2}}{2}\int_{0}^{t}\frac {d\sigma}{M(\sigma)}\} 
\exp\{-i\int_{0}^{t}G(\sigma)d\sigma+ix\beta(t)\}
\end{equation}
with $\beta(t)=-\int_{0}^{t} g_{1}(\sigma)d\sigma$ and $\nu(t)=-\int_{0}^{t}\frac {\beta(\sigma)}{M(\sigma)}d\sigma$.
To compare with the solution in Ref.[14], we let $g_{1}(t)$ take the form of $q(\epsilon_{0}+\epsilon\cos\omega t)$, and 
set $M(t)=m$, which yields
$$\Psi (x,t) = \frac {1}{\sqrt{2\pi}} \exp \{iA[x+\frac {q}{m}(\frac {\epsilon_{0}}{2}t^{2}-\frac {\epsilon}{\omega^{2}}
\cos\omega t+\frac {\epsilon}{\omega^{2}})]-i\frac {A^{2}t}{2m}\} \exp \{-i\frac {q}{\omega}(\epsilon_{0}\omega t
+\epsilon\sin \omega t)x\} $$
\begin{equation}
\exp \{-i\frac {q^{2}}{2m\omega^{3}}[\frac {\epsilon^{2}_{0}(\omega t)^{3}}{3} +
2\epsilon_{0}\epsilon (\sin\omega t-\omega t\cos\omega t)+\epsilon^{2}(\frac {1}{2}\omega t-\frac {1}{4}\sin 2\omega t)]\}.
\end{equation}
Obviously, when $A=0$, our solution is equivalent to Eq.(18) of Ref.[14] ${[18]}$.
As the physical meaning of $A$ is the momentum component of the free particle along the propagating direction, the solution in 
Ref.[14] can be considered as a special case that the particle is 'standing' in the potential of $g_{1}(t)x$.

In fact, for Eq.(5), besides the solution with the wave function of the plane wave, there is a remarkable but not widely known solution, 
called 'nonspreading wave packet' or 'Airy packet' solution 
with the form of $Ai[B(y-\frac {B^{3}s^{2}}{4})]\exp \{i\frac {B^{3}s}{2}(y-\frac {B^{3}s^{2}}{6})\}$, in which $B$ is an arbitrary
constant and $Ai$ the Airy function ${[19]}$. So the wave function of Eq.(1) is 
$$\Psi(x,t)=Ai[B(x+\int_{0}^{t}\frac {d\tau}{M(\tau)}\int_{0}^{\tau}g_{1}(\sigma)d\sigma -\frac {B^{3}}{4}(\int_{0}^{t}\frac {d\sigma}{M(\sigma)})^{2})]$$
$$\exp \{i\frac {B^{3}}{2}\int_{0}^{t}\frac {d\sigma}{M(\sigma)}[x+\int_{0}^{t}\frac {d\tau}{M(\tau)}\int_{0}^{\tau}g_{1}(\sigma)d\sigma 
-\frac {B^{3}}{6}(\int_{0}^{t}\frac {d\sigma}{M(\sigma)})^{2}]\} $$
\begin{equation}
\exp \{ -\frac {i}{2}\int_{0}^{t}\frac {d\tau}{M(\tau)} [\int_{0}^{\tau}g_{1}(\sigma)d\sigma]^{2}\}
\exp \{-ix\int_{0}^{t}g_{1}(\sigma)d\sigma\}.
\end{equation}
If we set $M(t)=m$, we will find that Eq.(8) here is formally different from Eq.(15) in Ref.[15] $[19]$. However, the correctness 
of Eq.(8) can be tested simply by substituting Eq.(8) into Eq.(1) and using the Airy function's properties${[20]}$.
It is easily seen that the probability density $|\Psi(x,t)|^{2}$ moves without change of form. The  Airy packet 
propagates along the trajectory given by
\begin{equation}
x_{0}(t)= \frac {B^{3}}{4}[\int_{0}^{t}\frac {d\sigma}{M(\sigma)}]^{2}-\int_{0}^{t}\frac {1}{M(\tau)}\int_{0}^{\tau}g_{1}(\sigma)d\sigma .
\end{equation}
If we choose $M(t)=m$ and $g_{1}(t)=\frac {B^{3}}{2m}$, then $x_{0}(t)=0$, which means the Airy packet is at rest. 
More specifically, setting $g_{1}(t)=q(\epsilon_{0}+\epsilon\cos\omega t)$ will yield 
$$\Psi(x,t)= Ai[B(x+ \frac {q\epsilon_{0}}{2m}t^{2}-\frac {q\epsilon}{m\omega^{2}}\cos\omega t 
+\frac {q\epsilon}{m\omega^{2}}-\frac {B^{3}t^{2}}{4m^{2}})]$$
$$\exp \{i\frac {B^{3}t}{2m}[x+ \frac {q\epsilon_{0}}{2m}t^{2}-\frac {q\epsilon}{m\omega^{2}}\cos\omega t 
+\frac {q\epsilon}{m\omega^{2}} -\frac {B^{3}t^{2}}{6m^{2}}]\}\exp \{-ix(q\epsilon_{0}t+\frac {q\epsilon}{\omega}\sin\omega t)\} $$
\begin{equation}
\exp\{-i\frac {q^{2}}{2m}[\frac {\epsilon_{0}^{2}}{3}t^{3}+\frac {\epsilon^{2}}{2\omega^{3}}(\omega t-\frac {1}{2}\sin 2\omega t)
+\frac {2\epsilon_{0}\epsilon}{\omega^{3}}(\sin\omega t -t\omega\cos\omega t)]\}.
\end{equation}
In this case the Airy packet propagates sinusoidally with no change of form.

In summary, we have presented the completely analytical solution of a system with a time-dependent-mass particle moving
in a time-dependent linear potential. 
The solution based on the wave function of the plane wave shows that the result in Ref.[14] is merely the solution of the system
in a particular case. The other solution with the form 
of the Airy function gives the specific form of a solution of the system, which has not been presented before. 
Each of the solutions has a free parameter representing the velocity of the wave packet of the solution.
As it is exact and complete, the present result can be used to investigate quantum properties of the system with a time-dependent linear 
potential in a wider range of parameters, and serve as a comparison with other approximate works.
We hope that the present work would be helpful for the future exploration in this respect.

The discussion with I.Guedes and Hanting Wang is highly acknowledged. The work is partly supported by the National Natural Science
Foundation of China.

{\bf Note added}: After finishing this work, the author was informed that a former work [21] for an electron moving in a general time-dependent
electromagnetic field, whose result is very similar to that in the present work, had been carried out by generally constructing the
evolution operators. The solution in [21] based on the wave function of the plane wave is the same as in the present work, whereas their 
solution with the Airy function is only for a special value of $B$ defined in Eq.(10).


\begin{thebibliography}{22}
\bibitem{18} There is a printing error in Eq.(18) of Ref.[14], i.e., $\frac {\epsilon_{0}(\omega t)^{3}}{3}$ should be replaced 
by $\frac {\epsilon_{0}^{2}(\omega t)^{3}}{3}$.
\bibitem{19} If a printing error in Eq.(15) of Ref.[15] is corrected, i.e., replacing $\frac {B^{2}t^{2}}{4m^{2}}$ with 
$\frac {B^{3}t^{2}}{4m^{2}}$, we can easily find that Eq.(15) of Ref.[15] is actually identical to Eq.(8) in the present paper by simply
supposing $g_{1}(t)$ to be a specific function, for example, $g_{1}(t)=q(\epsilon_{0}+\epsilon\cos\omega t)$.
\end{thebibliography}
\end{document}